\documentclass[superscriptaddress,onecolumn,aps,prd,11pt,nofootinbib,preprintnumbers]{revtex4-2}

\usepackage{amsmath} 
\usepackage{amssymb}
\usepackage{graphicx}
\usepackage[pgf]{adjustbox}
\usepackage[normalem]{ulem}
\usepackage{afterpage}
\usepackage{lmodern}
\usepackage[colorlinks=true, pdfstartview=FitV, linkcolor=red, citecolor=blue, urlcolor=blue]{hyperref}

\newcommand{\im}{{\rm i}}
\newcommand{\e}{{\rm e}}
\newcommand{\rmd}{\mathrm{d}}
\newcommand{\bx}{\boldsymbol{x}}
\newcommand{\bp}{\boldsymbol{p}}

\newcommand{\rout}{\bgroup \color{red} \ULdepth=-.5ex \ULset}
\newcommand{\bout}{\bgroup \color{blue} \ULdepth=-.5ex \ULset}
\newcommand{\vout}{\bgroup \color{violet} \ULdepth=-.5ex \ULset}
\begin{document}

\title{FRG analysis for a relativistic BEC in arbitrary spatial dimensions}

\author{Fumio~Terazaki}
\affiliation{Department of Physics, Tokyo University of Science, Tokyo 162-8601, Japan}
\author{Kazuya~Mameda}
\affiliation{Department of Physics, Tokyo University of Science, Tokyo 162-8601, Japan}
\affiliation{RIKEN iTHEMS, RIKEN, Wako 351-0198, Japan}

\preprint{RIKEN-iTHEMS-Report-25}

\begin{abstract}%
A relativistic Bose–Einstein condensate (BEC) is studied within the complex scalar field theory using the functional renormalization group (FRG) under the local potential approximation.
We investigate fluctuation effects on the relativistic BEC through numerical analyses for various spatial dimensions and chemical potentials.
Our numerical results are consistent with the Mermin-Wagner theorem, and this consistency is also analytically confirmed from the flow equation.
We also discuss a numerical instability of the FRG in lower spatial dimensions, which is evadable for certain parameter choices.
\end{abstract}

\maketitle

%%%%%%%%%%%%%%%%%%%%%%%%%%%%%%%%%%%%%%%%%%%%%%%%
%%%%%%%%%%%%%%%%%%%%%%%%%%%%%%%%%%%%%%%%%%%%%%%%

\section{Introduction}
\label{sec:intro}

A Bose–Einstein condensate (BEC) gives rise to remarkable macroscopic quantum phenomena. 
Liquid helium, a nonrelativistic system, exhibits superfluidity characterized by frictionless flows~\cite{kapitza1938viscosity,allen1938flow} and quantized vortices~\cite{hall1956rotation}.
In relativistic contexts, pion BECs~\cite{Sawyer:1972cq,Barshay:1973ju,Baym:1973zk} may influence the neutron-star equation of state and the cooling via neutrino emission~\cite{Shapiro:1983du}, though the occurrence of pion condensation inside neutron stars is still argued~\cite{Pethick:2017} both theoretically~\cite{Fore:2023gwv,Vijayan:2023qrt} and experimentally~\cite{Yasuda:2018den}.
It has also been proposed that a BEC could serve as a dark matter candidate in both relativistic~\cite{Urena-Lopez:2008vpl} and nonrelativistic~\cite{Boehmer:2007um} settings.

The functional renormalization group (FRG) provides a nonperturbative framework for obtaining an effective potential with quantum fluctuation effects taken into account.
This potential is determined from an exact equation, known as the Wetterich flow equation~\cite{Wetterich:1992yh}.
Various methods are available to solve the equation, such as the grid method, which tracks the full potential~\cite{Schaefer:2004en}, and the Taylor expansion method, which focuses on the local structure near the potential minimum~\cite{Tetradis:1993ts}.
The resulting potential encodes the thermodynamic behavior of the system, and its minimum determines the value of the order parameter such as the relativistic BEC.

An ordered phase is affected by thermal and quantum fluctuations.
The fluctuations are amplified as the spatial dimension $d$ decreases.
As a general consequence, continuous symmetries cannot be spontaneously broken at finite temperature and $d\leq2$, as stated by the Mermin–Wagner (MW) theorem~\cite{Mermin:1966fe,Hohenberg:1967zz}.
Reproducing the MW theorem is a requirement of a reliable theoretical method, including the FRG.
For example, it is shown that an FRG analysis of an $\mathrm{O}(N)\times Z_2$ model is consistent with the MW theorem~\cite{Hawashin:2024dpp}.
The theorem is also demonstrated by performing a fixed-point analysis of the $\mathrm{O}(N)$ model in Euclidean spacetime of arbitrary dimension~\cite{Defenu:2014jfa,Mati:2014xma}.
According to Ref.~\cite{Defenu:2014jfa}, a Taylor expansion of the potential around the origin fails to reproduce the MW theorem, whereas an expansion around the minimum succeeds.

The MW theorem seems to be reproduced in the FRG with finite chemical potential $\mu$, but there is a subtlety suggested by Refs.~\cite{Svanes:2010we,Palhares:2012fv,Terazaki:2024evv}.
Their FRG calculations indicate that at $d=3$, the relativistic BEC is enhanced as $\mu$ increases.
The same trend is expected even in lower $d$ since the structure of the flow equation is nearly unchanged in arbitrary $d$, evoking the inconsistency with the MW theorem.
Reproducing this universal theorem is one of the essential requirements for ensuring the reliability of the FRG in the imaginary-time formalism, which is believed to be a powerful nonperturbative approach for finite density systems~\cite{Floerchinger:2008jf,Drews:2016wpi,Otto:2019zjy,Fu:2019hdw}.

The aim of this paper is to verify whether the FRG meets the above crucial requirement within the simplest model.
That is, we analyze the FRG of the complex scalar field theory, which is equivalent to the $\mathrm{O}(2)$ model, at finite $\mu$ in lower $d$.
This work would also be complementary to lattice simulations of the $\mathrm{O}(2)$ model at finite $\mu$.
While such simulations avoid the sign problem~\cite{Prokofev:2001ddj,Endres:2006xu} and enable computation of the $T$-$\mu$ phase diagram for relativistic BEC~\cite{Gattringer:2012df}, the MW theorem within the lattice framework~\cite{Rindlisbacher:2020wvl} has not been fully investigated.

%%%%%%%%%%%%%%%%%%%%%%%%%%%%%%%%%%%%%%%%%%%%%%%%
%%%%%%%%%%%%%%%%%%%%%%%%%%%%%%%%%%%%%%%%%%%%%%%%

\section{FRG Formalism} 
\label{sec:formalism}
In this paper, we analyze the complex scalar field theory~\cite{Kapusta:1981aa}, which is, in the imaginary-time formalism, described by the following action:
\begin{equation}
S=\int_x\varphi^*\left[-(\partial_\tau-\mu)^2-\nabla^2+\bar{m}^2+\bar{\lambda}|\varphi|^2\right]\varphi ,
\label{eq:S}
\end{equation}
where the ($d+1$) spacetime integral is defined as $\int_x=\int_0^\beta\rmd\tau\int\rmd^d\bx$.
Here $\varphi=\varphi(x)$ is a complex scalar field with a mass $\bar{m}$ and a self-interaction coupling $\bar{\lambda}$.
The temperature is introduced as $T=\beta^{-1}$, and $\mu$ represents the chemical potential associated with the $\mathrm{U(1)}$ charge density.
Throughout this paper, we restrict our analysis to non-negative chemical potential, $\mu\geq0$. 

We now introduce the FRG formalism, which derives the effective action including nonperturbative quantum effects.
We denote $\Gamma_k$ as the scale-dependent effective action obtained by integrating out high-momentum modes above the scale $k$.
This effective action interpolates between $\Gamma_{k=\Lambda}=S$ at the ultraviolet scale $\Lambda$ and the quantum effective action $\Gamma_{k=0}$, and obeys the Wetterich flow equation~\cite{Wetterich:1992yh}:
\begin{equation}
\frac{\partial\Gamma_k}{\partial k}=\frac{1}{2}{\rm Tr}\left[\frac{\partial R_k}{\partial k}\frac{1}{\Gamma_k^{(2)}+R_k}\right],
\label{eq:Wetterich}
\end{equation}
where $\Gamma_k^{(2)}$ represents the two point vertex function computed from $\Gamma_k$:
\begin{equation}
\left[\Gamma_k^{(2)}(p,q)\right]_{ij}=\frac{\delta^2\Gamma_k}{\delta\Phi_i^\dagger(p)\Phi_j(q)},\quad
\Phi(p)=\begin{pmatrix}\varphi(p)\\ \varphi^*(-p)\end{pmatrix}.
\end{equation}
We denote the momentum variable as $p=(\omega_n,\bp)$ with the bosonic Matsubara frequency $\omega_n=2\pi n T$ and $\bp$ is the spatial momentum, and the trace ${\rm Tr}$ is taken with respect to the momentum indices $(p,q)$ and the field indices $i,j$.
We introduced the regulator function as $R_k(p,q)=\delta^{(d+1)}(p-q)R_k(p)$ with the $(d+1)$-dimensional delta function $\delta^{(d+1)}(p-q)=\beta\delta_{n_p,n_q}(2\pi)^d\delta^{(d)}(\bp-\boldsymbol{q})$.

While the flow equation~\eqref{eq:Wetterich} is exact, the practical analysis demands a further approximation to truncate an infinite hierarchy of coupled equations of the $n$-point vertex function $\Gamma_k^{(n)}$ for $n=0,1,2,\cdots$~\cite{Berges:2000ew}.
In this work, we adopt the derivative expansion~\cite{Golner:1985fg} together with the local potential approximation~\cite{Nicoll:1974zz}.
The scale dependent effective action is in general written as
\begin{equation}
\Gamma_k=\int_x\biggl[\varphi^*\Bigl(-(\partial_\tau-\mu)^2-\nabla^2\Bigr)\varphi+V_k(\rho)\biggr].
\label{eq:Gammak}
\end{equation}
An important remark is that a potential $V_k(\rho)$ with $\rho=|\varphi (x)|^2$ is introduced here;
the $\rho$ dependence is required by $\mathrm{U(1)}$-invariance of $\Gamma_k$.
This potential involves all higher-order terms of $|\varphi|$ generated by the flow of $\Gamma_k$, as in the Wilsonian renormalization group~\cite{Wilson:1973jj,Peskin:1995ev}.
When the ground state is translationally invariant, we can replace $\varphi(x)$ and $\varphi^*(x)$ with constants and reduce $\Gamma_k$ as $\Gamma_k = U_k(\rho) \times \int_x$ with the effective potential
\begin{equation}
U_k(\rho)=V_k(\rho)-\mu^2\rho.
\end{equation}

Further analysis of the FRG is performed with specifying the form of the regulator function $R_k$.
Let us first take the Litim regulator~\cite{Litim:2001up}
\begin{equation}
R_k(p)=(k^2-\bp^2)\theta(k^2-\bp^2),
\label{eq:Litim}
\end{equation}
which is advantageous in terms of numerical stability and efficiency.
Plugging this regulator into Eq.~\eqref{eq:Wetterich}, we derive the flow equation for $U_k$.
In the same manner as that for $d=3$ (see the Appendix in Ref.~\cite{Terazaki:2024evv}), we get
\begin{equation}
\label{eq:Uflow1}
\begin{split}
\frac{\partial U_k(\rho)}{\partial k}
&=
T\sum_{n=-\infty}^{\infty}\int\frac{\rmd^dp}{(2\pi)^d}\frac{\partial R_k(p)}{\partial k}\frac{\omega_n^2-\mu^2+\bp^2+R_k(p)+V_k'+\rho V_k''}{\displaystyle{\prod_{\alpha=\pm 1}}\Bigl[(\omega_n+\alpha\im\mu)^2+\bp^2+R_k(p)+V_k'+\rho V_k''\Bigr]-(\rho V_k'')^2}
\\
&=k^{d+1}a_d\sum_{s=\pm}\frac{E_s^2+\mu^2-E_k^2}{E_s^2-E_{-s}^2}\frac{1+2n_\mathrm{B}(E_s)}{E_s},
\end{split}
\end{equation}
where $n_\mathrm{B}(E)=(\e^{\beta E}-1)^{-1}$ and
\begin{equation}
    \begin{split}
        E_k^2=k^2+\mu^2+U_k'+\rho U_k'', \quad
        E_\pm=\sqrt{E_k^2+\mu^2\pm\sqrt{4\mu^2E_k^2+(\rho U_k'')^2}} ,
    \end{split}
\end{equation}
and the prime symbol denotes the derivative with respect to $\rho$.
Also the $d$-dimensional momentum integral yields the $d$-dependent numerical factor
\begin{equation}
 a_d=\frac{(4\pi)^{-d/2}}{\Gamma(1+d/2)}
\end{equation}
with the gamma function $\Gamma(z)$. 
Although different notation is used, the same flow equation is derived at $d=3$ in Ref.~\cite{Palhares:2012fv}.

For low $d$, the numerical computation of the flow equation~\eqref{eq:Uflow1} requires a high cost unlike for $d=3$.
In order to make it more tamable, we expand $U_k$ in terms of $\rho$ as
\begin{equation}
U_k(\rho)=\sum_{l=0}^{l_{\rm max}}\frac{u_{l,k}}{l!}(\rho-\rho_{0,k})^l,
\quad
\rho_{0,k}=\underset{\rho}{{\rm argmin}}\,U_k(\rho),
\label{eq:taylor}
\end{equation}
where we define $l_\mathrm{max}$ is the maximum order of the expansion.
The potential minimum $\rho_{0,k}$ corresponds to the squared value of the scale dependent relativistic BEC.
Differentiating Eq.~\eqref{eq:taylor} with respect to $k$ and $\rho$, we obtain the flow equations of the expansion coefficients $u_{l,k}$ as
\begin{equation}
\label{eq:usflow}
\frac{\partial^l}{\partial\rho^l}\left.\frac{\partial U_k}{\partial k}\right|_{\rho=\rho_{0,k}}
=\frac{\partial u_{l,k}}{\partial k}-u_{l+1,k}\frac{\partial\rho_{0,k}}{\partial k},
\qquad
l=0,1,...,l_{\rm max},
\end{equation}
where we impose $u_{l_{\rm max}+1,k}=0$.
The system of equations~\eqref{eq:usflow} can be numerically solved together with Eq.~\eqref{eq:Uflow1}.
Then, the initial condition at $k=\Lambda$ is provided by $\Gamma_{k=\Lambda}=S$, which implies $U_{k=\Lambda}(\rho) = (\bar{m}^2 - \mu^2)\rho + \bar{\lambda} \rho^2$, namely, at $\mu^2\geq\bar{m}^2$,
\begin{equation}
u_{0,k=\Lambda}=-\frac{(\mu^2-\bar{m}^2)^2}{4\bar{\lambda}},
\quad
u_{1,k=\Lambda}=0,
\quad
u_{2,k=\Lambda}=2\bar{\lambda},
\quad
\rho_{0,k=\Lambda}=\frac{\mu^2-\bar{m}^2}{2\bar{\lambda}},
\label{eq:initial}
\end{equation}
and at $\mu^2<\bar{m}^2$,
\begin{equation}
u_{0,k=\Lambda}=0,
\quad
u_{1,k=\Lambda}=\bar{m}^2-\mu^2,
\quad
u_{2,k=\Lambda}=2\bar{\lambda},
\quad
\rho_{0,k=\Lambda}=0.
\end{equation}
In both cases, $u_{l,k=\Lambda} = 0$ for $l>2$.
All dimensionful parameters $T$, $\mu$, $\bar{m}$, and $\bar{\lambda}$ are taken to be much smaller than the ultraviolet scale~$\Lambda$.

There are two important remarks on numerical computation.
First, although Eq.~\eqref{eq:usflow} seems to be the system of $l_{\rm max}+1$ equations to determine $l_{\rm max}+2$ functions, this is not an underdetermined system.
This is because the nonzero minimum $\rho_{0,k}>0$ requires $u_{1,k}=\partial_\rho U_k|_{\rho=\rho_{0,k}}=0$ by definition, and for $u_{1,k}=\partial_\rho U_k|_{\rho=\rho_{0,k}}>0$ we always have $\rho_{0,k}=0$.

Another is about the numerical instability, which is caused by 
some parameter choices of $d$, $T$, $\mu$, $\bar{m}$, and $\bar{\lambda}$.
A good example to illustrate this problem is Eq.~\eqref{eq:usflow} for $\mu=0$ and $l=1$ as
\begin{equation}
\label{eq:rhoflow1}
 k\frac{\partial\rho_{0,k}}{\partial k}
 =\frac{a_d k^{d-1}}{4}\biggl[
 F(k) + (3+2y_k) \frac{F(k\sqrt{1+2x_k})}{(1+2x_k)^{3/2}}
 \biggr],
\end{equation}
where we introduce the following shorthand notations
$n_\mathrm{B}'(z) = \rmd n_\mathrm{B}(z)/\rmd z$, $F(z)=1+2n_\mathrm{B}(z)-2 z\, n_\mathrm{B}'(z)$ and
\begin{equation}
\label{eq:xy}
 x_k=\frac{\rho_{0,k}u_{2,k}}{k^2},
\qquad
y_k=\frac{\rho_{0,k}u_{3,k}}{u_{2,k}}.
\end{equation}
The flow equation~\eqref{eq:rhoflow1} converges as long as the right-hand side is regular in the limit of $k\to 0$.
A general feature of well-defined theories, $u_{2,k\to0}\geq 0$~\cite{Berges:2000ew}, implies $x_k=\rho_{0,k}u_{2,k}/k^2\geq 0$, for which $(1+2x_k)^{-3/2}F(k\sqrt{1+2x_k})$ is not singular.
In general, however, another $k$-dependent function $y_k$ can diverge more rapidly than $k^{-d+1}$.
As seen from the definition $y_k=\rho_{0,k} u_{3,k}/u_{2,k}$, this divergence occurs when the potential $U_k(\rho)$ becomes highly flat around the minimum.
The numerical computation in this paper is performed only for good parameters evading the numerical instability;
we will again argue this problem in the following sections.

We briefly comment on the applicability of the Taylor expansion.
In general, this method provides only a potential shape around a minimum $\rho_{0,k}$.
For this reason, if $\rho_{0,k-\Delta k}$ were largely different from $\rho_{0,k}$, the flow of the minimum would no longer be correctly tracked.
We have confirmed that such a troublesome behavior in the following numerical calculations is not observed from another numerical computation based on the Grid method, which is applicable even to the case with discontinuous changes of $\rho_{0,k}$.
This smooth flow of $\rho_{0,k}$ ensures the validity of our analysis based on the Taylor expansion.

%%%%%%%%%%%%%%%%%%%%%%%%%%%%%%%%%%%%%%%%%%%%%%%%
%%%%%%%%%%%%%%%%%%%%%%%%%%%%%%%%%%%%%%%%%%%%%%%%

\section{Numerical Results} 
\label{sec:numerical}

\begin{figure}
\centering
\begin{minipage}{0.48\textwidth}
    \centering
    \includegraphics[width=\linewidth]{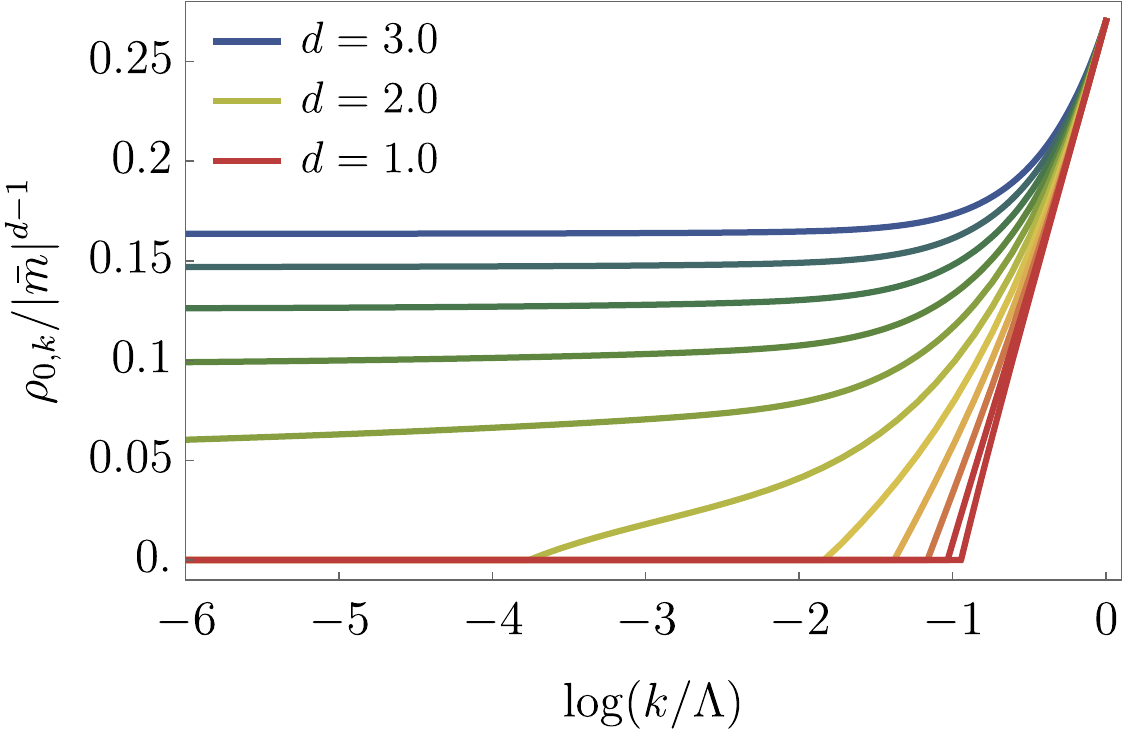}
\end{minipage}
\hfill
\begin{minipage}{0.48\textwidth}
    \centering
    \includegraphics[width=\linewidth]{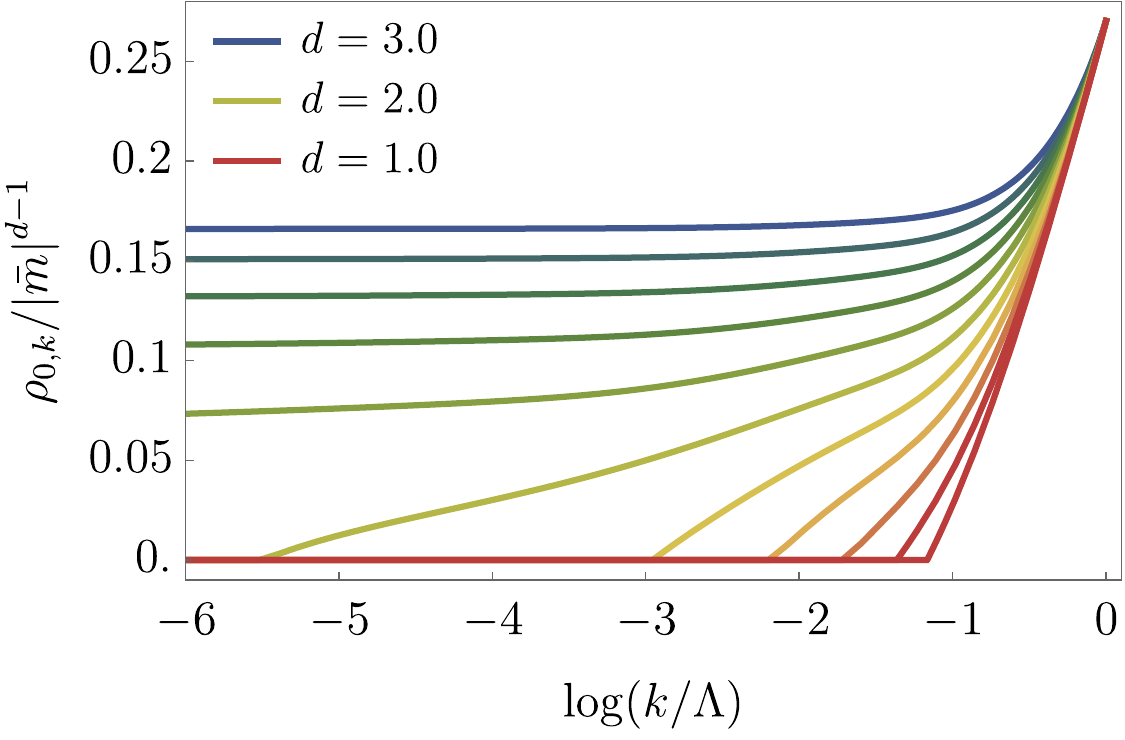}
\end{minipage}
    \caption{
    Flows of $\rho_{0,k}$ at $\mu=0$ (left) and $\mu>0$ (right).
    In the left panel, $\bar{\lambda}/|\bar{m}|^{3-d} = 37/20$; in the right panel, $\bar{\lambda}/|\bar{m}|^{3-d} = 1/15$ and $\mu/|\bar{m}| = \sqrt{115/111}$.
    In both panels, the spatial dimension $d$ varies from top to bottom as $3.0, 2.8, \dots, 1.0$.
    }
    \label{fig:flow}
\end{figure}

\begin{figure}
\centering
\begin{minipage}{0.48\textwidth}
    \centering
    \includegraphics[width=\linewidth]{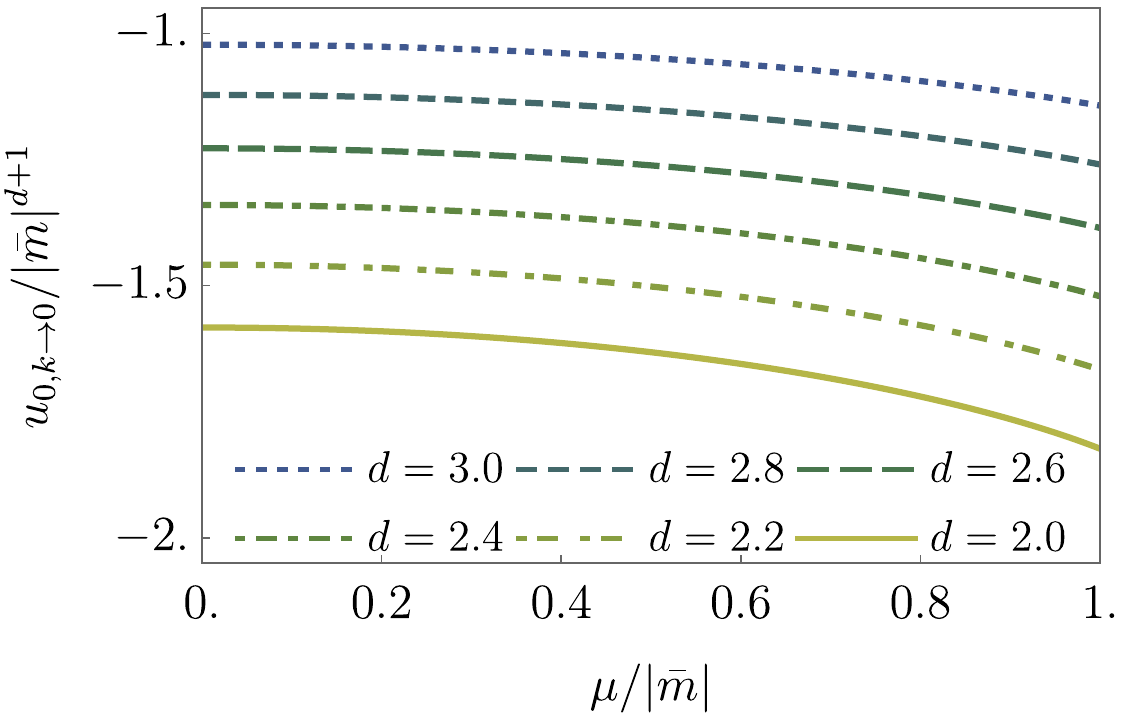}
\end{minipage}
\hfill
\begin{minipage}{0.48\textwidth}
    \centering
    \includegraphics[width=\linewidth]{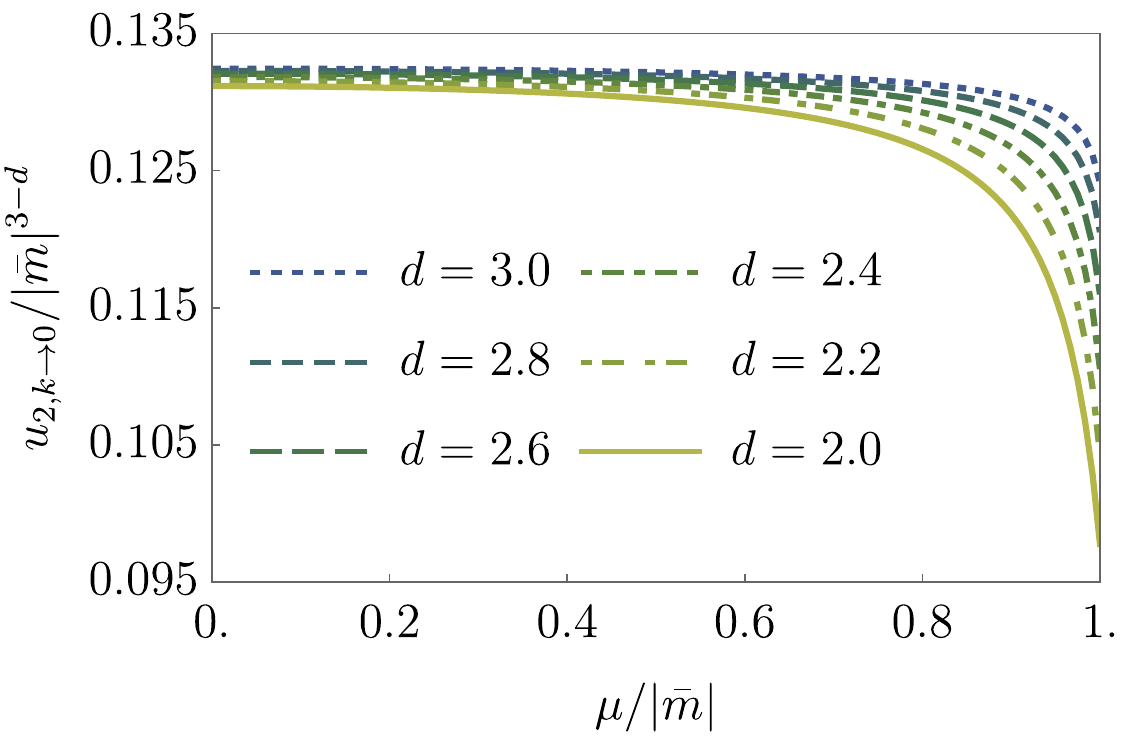}
\end{minipage}
\caption{
Chemical potential dependence of Taylor coefficients $u_{0,k\to0}$ (left) and $u_{2,k\to0}$ (right) at $T=0$.
The spatial dimension $d$ of each line from top to bottom corresponds to $d=3.0, 2.8, \dots, 2.0$.
}
\label{fig:SB}
\end{figure}

We now show numerical results for the above FRG formulation.
We fix $\bar{m}/\Lambda=0.25\im$ for $\mu=0$, and $\bar{m}/\Lambda=0.25$ for $\mu>0$.
Due to Eq.~\eqref{eq:initial}, these choices provide $\rho_{0,k=\Lambda}>0$, with which we can elucidate fluctuation effects on the relativistic BEC;
we have numerically confirmed that no fluctuation effect appears when we take $\rho_{0,k=\Lambda}=0$.
We also fix temperature as $T/|\bar{m}|=0.1$ (except in Fig.~\ref{fig:SB}, where $T=0$, and in calculations of critical exponents at the end of this section, where $T/|\bar{m}|=2$ or $0$), and set $\bar{\lambda}/|\bar{m}|^{3-d}=1/15$ at finite $\mu$ (except in the left panel of Fig.~\ref{fig:flow}, where $\mu=0$), and vary $\mu$ and $d$.
The effective potential $U_k$ is expanded up to the seventh order (i.e., $l_{\rm max} = 7$), for which the numerical results are convergent enough.

In Fig.~\ref{fig:flow}, we show the scale dependence of $\rho_{0,k}$.
In the left and right panels, we employ $(\mu/|\bar{m}|,\bar{\lambda}/|\bar{m}|^{3-d})=(0,37/20)$ and $(\mu/|\bar{m}|,\bar{\lambda}/|\bar{m}|^{3-d})=(\sqrt{115/111},1/15)$, respectively, both of which provide the same $\rho_{0,k=\Lambda}$ in Eq.~\eqref{eq:initial}.
Each line from top to bottom corresponds to $d=3-0.2j$ with $j=0,1,\cdots,10$, respectively.
It is obvious that the infrared value $\rho_{0,k\to 0}$ is always smaller than the ultraviolet one $\rho_{0,k=\Lambda}$.
This suppression is understandable as the effect of thermal and quantum fluctuations.

Figure~\ref{fig:flow} also exhibits the apparent difference between $d>2$ (i.e., the upper five lines) and $d\leq 2$ (i.e., the lower six lines).
For $d>2$, the condensate $\rho_{0,k}$ converges to a finite value of $\rho_{0,k\to0}$.
The convergence for $d=2.2$ is much slower, but our numerical calculation shows that the flow eventually converges to $\rho_{0,k\to 0}/|\bar{m}|^{d-1}\approx 0.049$ (\text{left}) and $\rho_{0,k\to 0}/|\bar{m}|^{d-1}\approx 0.062$ (\text{right}) around $\log(k/\Lambda)\approx -25$.
These trends for $d>2$ meet the intuitive picture that fluctuations become more diluted in higher spatial dimensions.
On the other hand, the plots for $d\leq 2$ exhibit the vanishing of the BEC, implying the restoration of $\mathrm{U(1)}$ symmetry.
The two different behaviors for $d>2$ and $d\leq 2$ are consistent with the MW theorem.

\begin{figure}
\centering
\begin{minipage}{0.48\textwidth}
    \centering
    \includegraphics[width=\linewidth]{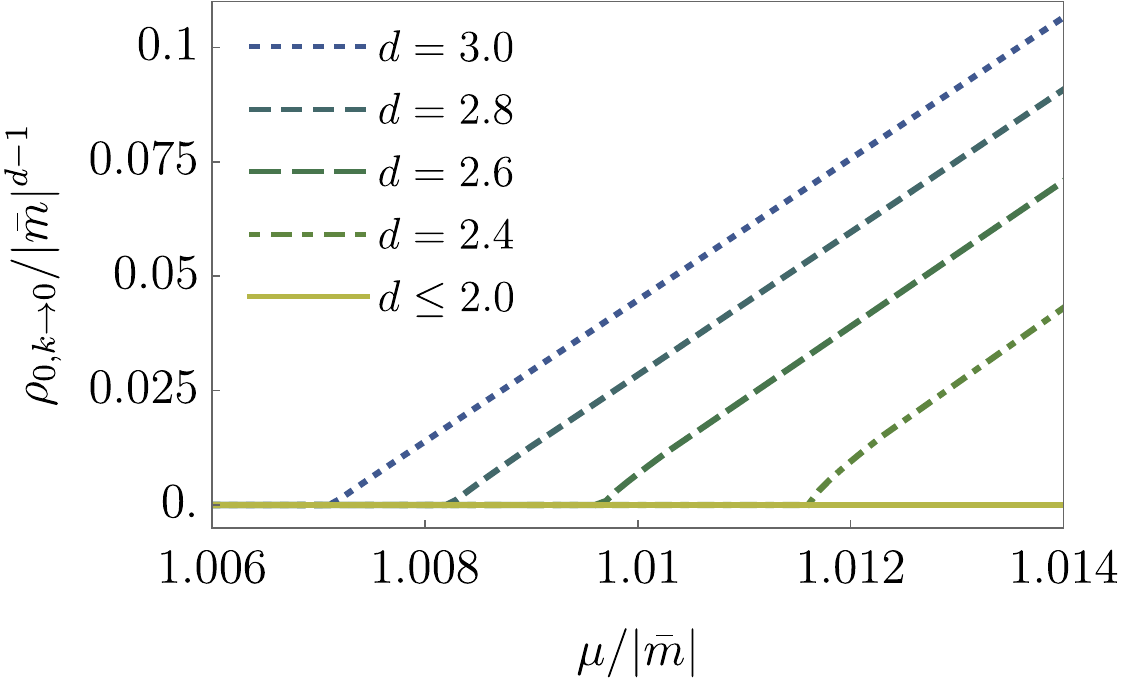}
    \caption{Dependence of $\mu$ on $\rho_{0,k\to 0}$.
    The spatial dimension $d$ for each rising line varies from top to bottom as $3.0, 2.8, 2.6$, and $2.4$. 
    The horizontal line corresponds to $d\leq2.0$.
    }
    \label{fig:rho0_mu}
\end{minipage}
\hfill
\begin{minipage}{0.48\textwidth}
    \centering
    \includegraphics[width=\linewidth]{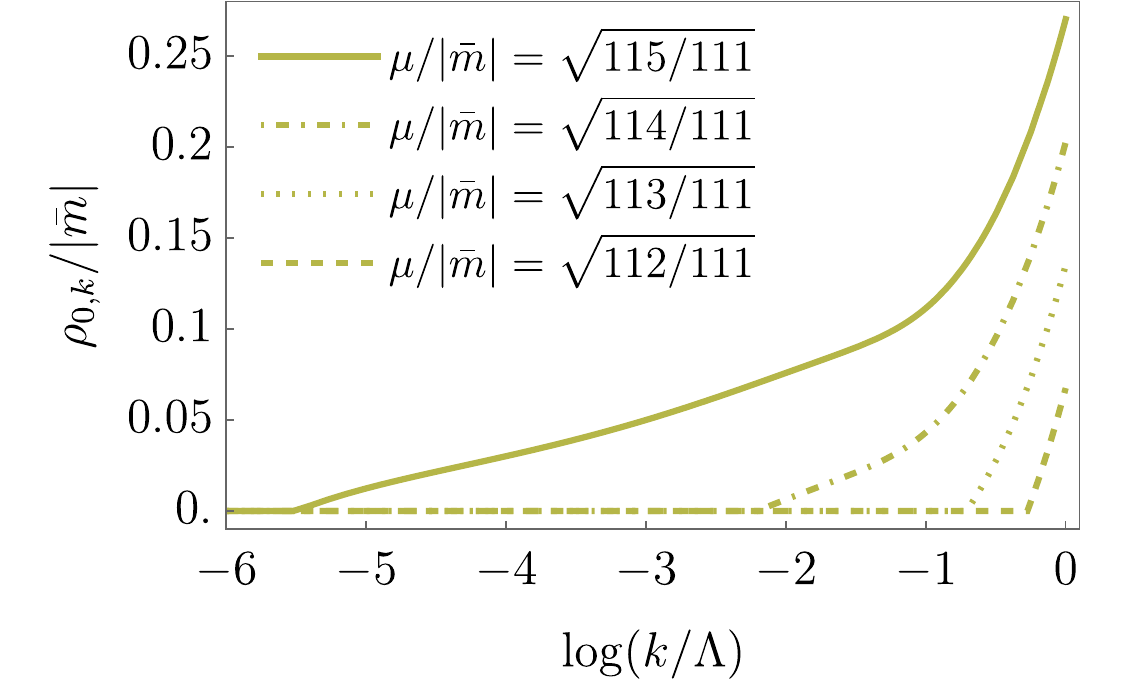}
    \caption{Flows for $d=2$ with various $\mu$.
    The chemical potential $\mu$ for each line, from top to bottom, is given by $\sqrt{\mu/|\bar{m}|}=\sqrt{(115 - j)/111}$, where $j=0,1,2,3$.
    }
    \label{fig:flow_mus}
\end{minipage}
\end{figure}

In Fig.~\ref{fig:rho0_mu}, we make a plot of the $\mu$ dependence of the condensate $\rho_{0,k\to 0}$.
Each line from top to bottom again corresponds to $d=3-0.2j$ with $j=0,1,2,3$, and $d\leq2$ respectively.
For higher dimensions $d>2$, we observe that the enhancement of $\rho_{0,k\to 0}$ by larger $\mu$, as in the case of $d=3$~\cite{Terazaki:2024evv,Svanes:2010we}.
The initial condition of the potential minimum~\eqref{eq:initial} clearly shows that the $\mathrm{U}(1)$ symmetry breaking becomes stronger at the UV scale as $\mu$ increases, favoring larger condensates.
Therefore, the increase in $\rho_{0,k\to0}$ with larger $\mu$ can be regarded as a natural consequence, at least from the change in the initial values.
In contrast, the condensate $\rho_{0,k\to 0}$ for $d\leq 2$ are totally irrelevant to $\mu$.
Hence, our FRG analysis demonstrates that the suppression by fluctuations 
for $d\leq 2$ overcomes the enhancement by a finite chemical potential while that for $d>2$ does not, leading to the consistency with the MW theorem.

Figure~\ref{fig:flow_mus} is a plot of the flow of the condensate $\rho_{0,k}$ for $d=2$ with different values of $\mu$.
Each plot from top to bottom corresponds to $\mu/|\bar{m}|=\sqrt{(115-j)/111}$ with $j=0,1,2,3$.
This figure exhibits that the condensate for $d=2$ vanishes independently of $\mu$, which again agrees with the MW theorem.
Another important observation is that the convergence becomes slower for larger $\mu$.
Such a behavior of slow convergence requires high precisions in numerical computation of the FRG.

\begin{figure}
    \centering
    \includegraphics[width=0.5\linewidth]{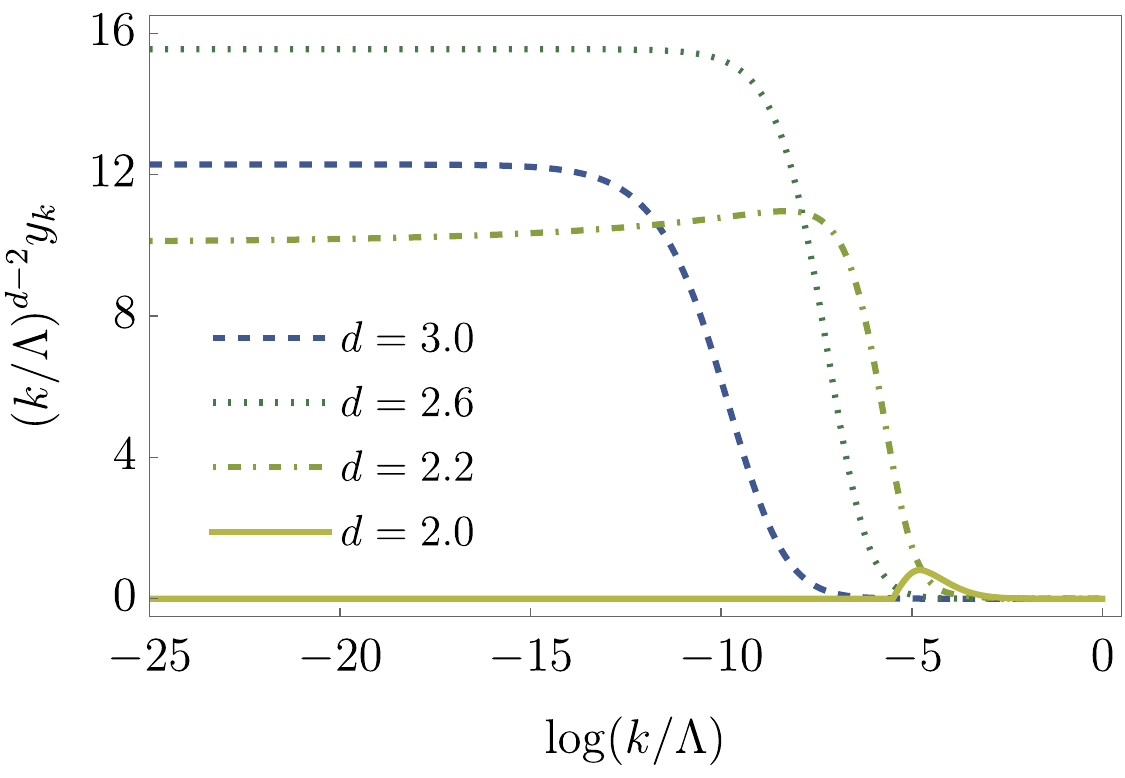}
    \caption{Flows of $(k/\Lambda)^{d-2}y_k$ with various $d$.
    The chemical potential is fixed at $\mu/|\bar{m}|=\sqrt{115/111}$.
    The spatial dimension $d$ corresponds to each line as follows: $d=3.0$ (blue dashed), $d=2.6$ (green dotted), $d=2.2$ (green dash-dotted), and $d=2.0$ (green solid).
    }
    \label{fig:yk}
\end{figure}

Figure~\ref{fig:yk} is a plot to illustrate the divergence behavior of $y_k$ with $\mu/|\bar{m}|=\sqrt{115/111}$.
The vertical axis is multiplied by $(k/\Lambda)^{d-2}$.
While the line for $d=2.0$ rapidly approaches to zero in the limit of $k \to 0$, other lines converge to constant values, implying that $y_k$ behaves like $\sim k^{-d+2}$.
This behavior indicates that slight modification of parameters potentially varies this power law of $y_k$ in the infrared region, leading to serious singularities in the system of equations~\eqref{eq:usflow}.
Indeed, we have checked that a slightly larger $\mu$, such as $\mu/|\bar{m}|=\sqrt{115.1/111}$, yields the numerical instability at $d=2$.
This divergence is not found for sufficiently larger $d$ thanks to the power-factor of $k$ involving the flow equation, such as $k^{d-1}$ in Eq.~\eqref{eq:rhoflow1}.

The problem above is a numerical artifact, which might be circumvented with higher numerical accuracy.
For instance, this artifact in general disappears by taking larger $l_\mathrm{max}$ in the Taylor expansion method.
Such a direct strategy is, however, impractical because Eq.~\eqref{eq:usflow} requires the computation of higher-order derivative terms, amplifying the numerical cost rapidly.
We note that the same problem can be circumvented by taking the large-$N$ limit of an $\mathrm{O}(N)$ model, where $O(\rho^l)$ contributions are suppressed by powers of $O(N^{-l})$~\cite{Hawashin:2024dpp}.
An alternative method to evade the instability could be the Grid method with higher accuracy.
Indeed, we have found some parameters for which the Taylor expansion method suffers from the instability, but the Grid method does not.
Nevertheless, the Grid method is again impractical to solve the instability problem due to $y_k=\rho_{0,k} u_{3,k}/u_{2,k}\to \infty$, which implies that a totally flat structure of the potential $U_k(\rho)$ is formed around the minimum $\rho_{0,k}$.
This flatness makes it difficult to accurately compute derivatives with respect to $\rho$.

In the rest of this section, we make several remarks on our analysis.
The first one is about the regulator dependence.
In the above numerical computations, we employ the Litim regulator~\eqref{eq:Litim}, which fulfills the optimization condition~\cite{Litim:2001up}.
As at finite temperature the Litim regulator yields reliable results, it is widely used in the imaginary-time formalism.
On the other hand, the Litim regulator violates the Lorentz invariance, leading to the inconsistency between the flow equations derived in the imaginary-time formalism in $T\to0$ limit and those formulated in Euclidean spacetime.
This fact implies, for instance, that the present analysis does not reproduce the Silver-Blaze property~\cite{Cohen:2003kd,Akerlund:2014mea,Marko:2014hea,Khan:2015puu}.
In order to illustrate this fact, in Fig.~\ref{fig:SB} we make a plot of the $\mu$ dependence of $u_{0,k\to 0}$ and $u_{2,k\to 0}$.
It is obvious that these Taylor coefficients depend on $\mu$ even at smaller $\mu \ll |\bar{m}|$.
While the Silver-Blaze property could be reproduced by using another appropriate regulator~\cite{Topfel:2024iop}, we leave it for future work.

\begin{figure}
\centering
\begin{minipage}{0.48\textwidth}
    \centering
    \includegraphics[width=\linewidth]{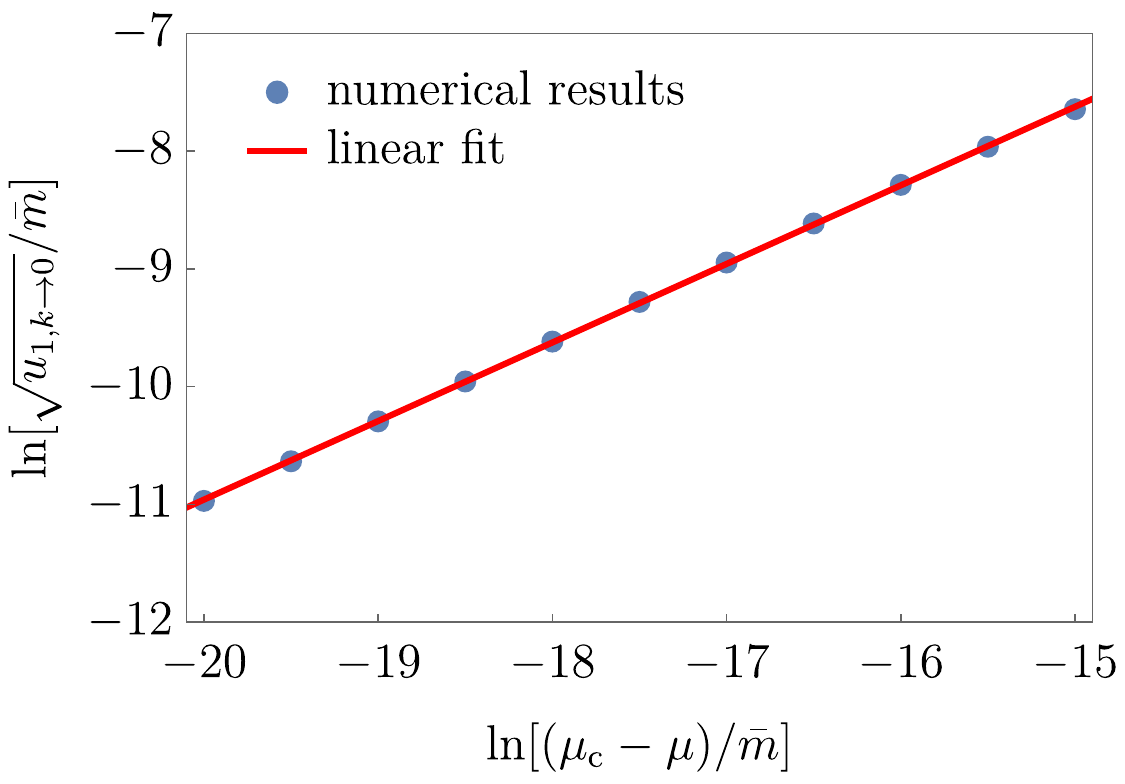}
\end{minipage}
\hfill
\begin{minipage}{0.48\textwidth}
    \centering
    \includegraphics[width=\linewidth]{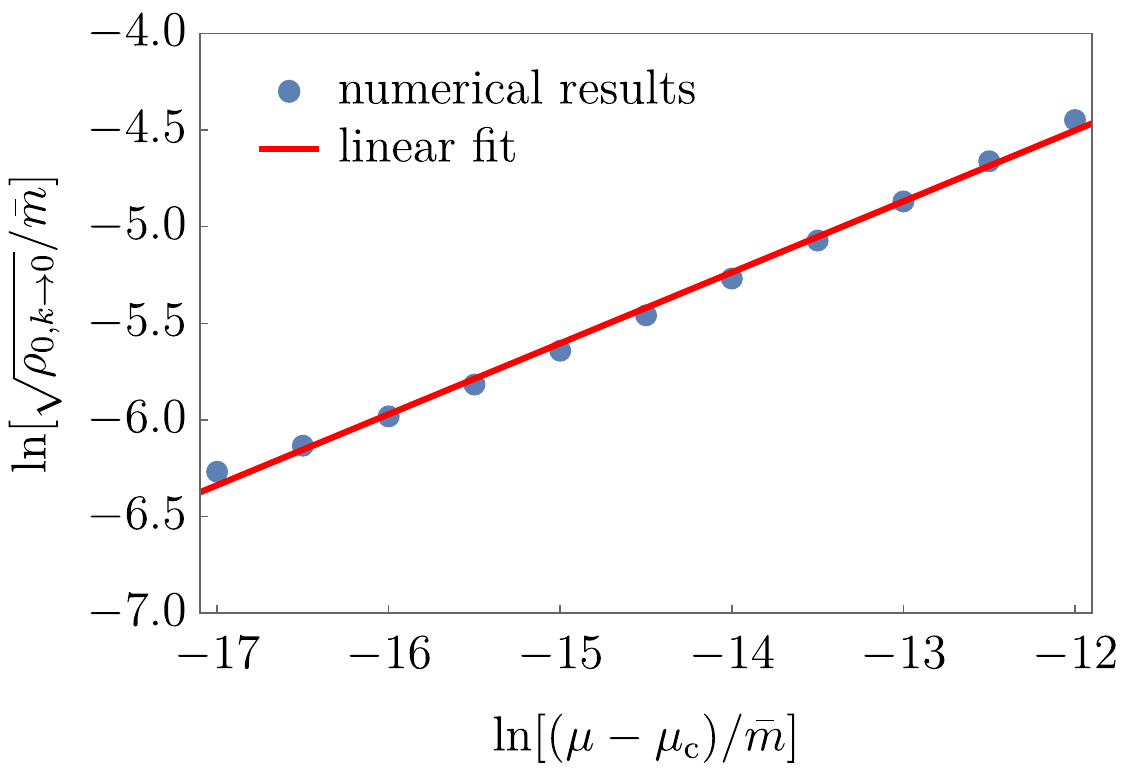}
\end{minipage}
\caption{
Critical exponents $\nu$ (left) and $\beta$ (right) at $d=3$.
The value of $\mu_{\rm c}/|\bar{m}|$ at $T/|\bar{m}|=0.1$ is obtained from the blue dotted line in Fig.~\ref{fig:rho0_mu}.
We perform the linear fitting (red line) to the numerical results (blue dots).
}
\label{fig:exponents}
\end{figure}

Another remark is that the present theory exhibits a second-order phase transition.
If such a phase transition takes place, the inverse correlation length (i.e., the mass of the complex scalar field) and the condensate behave as $\sqrt{u_{1,k\to0}}\sim(\mu_{\rm c}-\mu)^\nu$ for $\mu<\mu_{\rm c}$ and $\sqrt{\rho_{0,k\to0}}\sim(\mu-\mu_{\rm c})^\beta$ for $\mu>\mu_{\rm c}$, respectively~\cite{Bohr:2000gp,Akerlund:2014mea,Springer:2015kxa}.
The value of $\mu_{\rm c}$ is obtained from the blue dotted line ($d=3$) in Fig.~\ref{fig:rho0_mu}.
In Fig.~\ref{fig:exponents}, we confirm such power-law behaviors.
From the numerical fittings (red lines) for the numerical results (blue dots), we obtain the critical exponents $\nu\approx0.6672$ and $\beta\approx0.3670$.
This result ensures that our formalism based on the Taylor expansion method~\eqref{eq:taylor} is valid for analyzing phase transitions at finite $\mu$.

Finally, we make a comparison between the critical exponents in this study and those in preceding works.
While at high temperature the $(3+1)$-dimensional $\mathrm{O}(2)$ model belongs to the universality class of the three-dimensional XY model, at zero temperature its critical behavior becomes the same as the one in the four-dimensional free theory~\cite{Wang:2015bky}.
In the former and the latter, the critical exponents are evaluated as $(\nu,\beta)\approx (0.6717,0.3486)$~\cite{Hasenbusch:2019jkj}, and $(\nu,\beta)=(0.5,0.5)$~\cite{Akerlund:2014mea,Wang:2015bky}, respectively.
In similar calculations to Fig.~\ref{fig:exponents}, we numerically obtain $(\nu,\beta)\approx(0.7157,0.3502)$ at high temperature $T/|\bar{m}|=2$ and $(\nu,\beta)\approx(0.4980,0.4959)$ at $T=0$, which are in good consistent with those in the above preceding works.
We note that the deviation in the high temperature case would be attributed to the local potential approximation, in which anomalous dimension is set to zero.

%%%%%%%%%%%%%%%%%%%%%%%%%%%%%%%%%%%%%%%%%%%%%%%%
%%%%%%%%%%%%%%%%%%%%%%%%%%%%%%%%%%%%%%%%%%%%%%%%

\section{Analytical revisit}
\label{sec:linear}

While in Sec.~\ref{sec:numerical} we numerically confirmed the MW theorem at both $\mu=0$ and $\mu>0$, this section is devoted to the analytical revisit to the MW theorem as a crosscheck.
Instead of the Litim regulator~\eqref{eq:Litim}, here we employ the following regulator~\cite{Hawashin:2024dpp}:
\begin{equation}
R_k(p)=(k^2-\omega_n^2-\bp^2)\theta(k^2-\omega_n^2-\bp^2),
\label{eq:linear}
\end{equation}
which allows a more transparent analytical treatment of the flow equation in a low-momentum regime.
Through this approach, we can clarify the reason why the FRG prohibits the spontaneous symmetry breaking for $d\leq2$.

Let us first look at the flow equation of the potential $U_k=U_k(\rho)$.
Plugging the regulator~\eqref{eq:linear} into Eq.~\eqref{eq:Wetterich} and performing the $d$-dimensional momentum integration, we get
\begin{equation}
\frac{\partial U_k}{\partial k}
=\frac{k^{d+1}a_dT}{2\mu^2}
\sum_{n=-\kappa}^\kappa\frac{k^2+U_k'+\rho U_k''}{\omega_n^2+\tilde{E}^2}\left(1-\frac{\omega_n^2}{k^2}\right)^{d/2},
\label{eq:Uflow2}
\end{equation}
with $\tilde{E}=\sqrt{(k^2+U_k')(k^2+U_k'+2\rho U_k'')}/(2\mu)$ and the prime denote the $\rho$ derivative.
The sum of the Matsubara frequency is, unlike Eq.~\eqref{eq:Uflow1}, the finite sum bounded by $\kappa=\lfloor k/(2\pi T)\rfloor$, where $\lfloor x\rfloor$ is the floor function.
A crucial feature of this flow equation is the existence of the factor $(1-\omega_n^2/k^2)^{d/2}$, which is in general nonanalytical unless $d/2$ is an integer.
Due to this nonanalyticity, the bounded Matsubara sum cannot be represented by a contour integral.
For $d\neq 2$, the Matsubara summation at every step of $k$ runs up a numerical cost.
Nevertheless, for $d=2$, the flow equation~\eqref{eq:Uflow2} is much reduced to a tractable form as
\begin{equation}
\begin{split}
\frac{\partial U_k}{\partial k}
&=\frac{k(k^2+U_k'+\rho U_k'')T}{8\pi\mu^2}
\\&\quad
\times\left[
-2\kappa+\frac{k^2}{\tilde{E}^2}
+\im\frac{k^2+\tilde{E}^2}{2\pi T\tilde{E}}\sum_{\alpha=\pm1}\alpha\left\{\psi\left(1+\frac{\im \alpha\tilde{E}}{2\pi T}+\kappa\right)-\psi\left(1+\frac{\im\alpha\tilde{E}}{2\pi T}\right)\right\}
\right],
\end{split}
\label{eq:Uflow3}
\end{equation}
where we use
\begin{equation}
\sum_{n=1}^N\frac{1}{z+n-1}=\psi(z+N)-\psi(z) \quad
\text{for} \quad z\neq0,-1,-2,\cdots
\end{equation}
with the digamma function $\psi(z)=d\log \Gamma(z)/dz$.
We have numerically verified that Eq.~\eqref{eq:usflow} with Eq.~\eqref{eq:Uflow3} yields qualitatively the same results as those in Fig.~\ref{fig:flow_mus}.
For this reason, Eq.~\eqref{eq:Uflow2} works for the revisit to the results obtained with the Litim regulator~\eqref{eq:Litim}.

Let us focus on the low energy regime $k<2\pi T$, which is sufficient for the analysis of the phase transition.
Then, the Matsubara frequencies with $n\neq 0$ do not contribute to Eq.~\eqref{eq:Uflow2}.
Plugging the Taylor-expanded potential~\eqref{eq:taylor} and Eq.~\eqref{eq:Uflow2} into the recursion relation~\eqref{eq:usflow} for $l=1$, we arrive at the following flow equation of $\rho_{0,k}$:
\begin{equation}
k\frac{\partial\rho_{0,k}}{\partial k}
= 4 a_d T k^{d-2}f_k,\quad
f_k
=\frac{1+x_k+x_k^2+y_k/2}{(1+2x_k)^2},
\label{eq:rhoflow2}
\end{equation}
where $x_k$ and $y_k$ are defined in Eq.~\eqref{eq:xy}.
We note that while the explicit $\mu$ dependence disappears from this flow equation, the implicit $\mu$ dependence remains through the flow of $U_k(\rho)$ for $k\geq 2\pi T$, which is dependent on $\mu$.

While fully understanding the infrared behavior of $\rho_{0,k}$ needs numerical analysis of  
the flow equation~\eqref{eq:rhoflow2}, several features are extracted from analytical examination, as in Ref.~\cite{Hawashin:2024dpp}.
Let us first consider the specific case where we impose the following two assumptions:
One is that $y_k$ can be Taylor-expanded for a small $k$, namely, $f_k \simeq f_0 + O(k)$ with $f_0 = f_{k\to 0}$, and the other is $u_{3,k}\geq0$.
The latter assumption holds at least our numerical computation yielding Figs.~\ref{fig:flow}-\ref{fig:yk}, while the former is fulfilled by some parameter sets.
Since $f_{0}$ is independent on $k$, we can analytically solve Eq.~\eqref{eq:rhoflow2} as 
\begin{equation}
\rho_{0,k}\simeq
\begin{cases}
%\rho_{0,k_\mathrm{i}}+A^{-1}(k^{d-2}-k_\mathrm{i}^{d-2})/(d-2), & d\neq2,\\
\rho_{0,k_\mathrm{i}}-\Bigl[(-d+2)A\Bigr]^{-1}(k^{d-2}-k_\mathrm{i}^{d-2}), & d\neq2,\\
\rho_{0,k_\mathrm{i}}+A^{-1}\log(k/k_\mathrm{i}), & d=2,
\end{cases}
\end{equation}
where we define $A= (4Ta_d f_0)^{-1}$ and $k_\mathrm{i}$ is a scale that fulfills $k<k_\mathrm{i} \ll T,\mu,|\bar{m}|,\bar\lambda^{1/(3-d)}$.
Since $f_{0}$ is also positive as long as $u_{3,k}\geq0$ holds, for $d>2$ the condensate $\rho_{0,k}$ does not necessarily vanish.
On the other hand, for $d\leq 2$ there exist a value of $k$ at which $\rho_{0,k} = 0$.
For $d\leq 2$, the critical scale below which the condensate vanishes is derived as follows 
\begin{equation}
\label{eq:kc}
k_\mathrm{c}\simeq
\begin{cases}
\displaystyle k_\mathrm{i} \left[1+(2-d) k_\mathrm{i}^{2-d}\rho_{0,k_\mathrm{i}} A\right]^{-1/(2-d)}, & d<2, 
\\
\displaystyle k_\mathrm{i}\exp(-\rho_{0,k_\mathrm{i}}A), & d=2.
\end{cases}
\end{equation}
Since the assumption $u_{3,k}\geq0$ leads to $\partial\rho_{0,k}/\partial k>0$, for $d\leq2$ arbitrary low scales $k\leq k_\mathrm{c}$ provide $\rho_{0,k}=0$.
This completes the confirmation of the MW theorem under the assumption that $u_{3,k}\geq0$ and $f_k$ can be Taylor-expanded.
This analytical result partially explains the reason why the numerical results shown in Figs.~\ref{fig:flow}-\ref{fig:flow_mus} is consistent with the MW theorem.

Although the above argument is inapplicable to the case where $f_k$ is singular at the infrared regime, a similar analysis provides a different perspective of the numerical instability.
As a simple example, we suppose that in the limit of $k\to 0$, the function $f_k$ takes the form of $f_k=\tilde{f}_0\,k^{-p}$ with a $k$-independent factor $\tilde{f}_0$ and $p>0$. 
For $p\geq d-2$, the critical scale $k_\mathrm{c}$ takes a similar form to Eq.~\eqref{eq:kc} but with the replacement $d\to d-p$ and $A\to (4Ta_d \tilde{f}_0)^{-1}$, while the case for a smaller $p<d-2$ does not necessarily have a solution $k_\mathrm{c}$.
Since $p=d-2$ is a threshold where the algebraic structure sharply changes, we need highly accuracy to determine the value of $p$;
otherwise, a slightly different $p$ could provide much different flows of $\rho_{0,k}$.
This is an additional issue coming from the singular structure of $f_k$ or $y_k$, on top of the numerical instability described in the previous sections.

%%%%%%%%%%%%%%%%%%%%%%%%%%%%%%%%%%%%%%%%%%%%%%%%
%%%%%%%%%%%%%%%%%%%%%%%%%%%%%%%%%%%%%%%%%%%%%%%%

\section{Summary}
\label{sec:summary}

In this work, we have studied the relativistic Bose–Einstein condensate (BEC) of the complex scalar field theory, based on the functional renormalization group (FRG) in the imaginary-time formalism.
For numerical analysis, we have employed the Litim regulator together with the local potential approximation.
We have numerically analyzed the flow equation of the scale dependent relativistic BEC with the Taylor expansion at various spatial dimensions $d$ and chemical potential $\mu$.
The results indicate that fluctuations tend to suppress the BEC, suggesting that fluctuation effects become weaker as $d$ increases due to their dilution across more directions.
Although for $d>2$ larger $\mu$ enhances the condensate, for $d\leq2$ the condensate converges to zero due to the effect of fluctuations exceeding that of $\mu$, in accordance with the Mermin–Wagner (MW) theorem.
We have found that the flow equation becomes numerically unstable in some parameter regions, for example, low $d$ and high $\mu$.
In this work, employing the Litim regulator, we examine the FRG in the imaginary-time formalism.
As this regulator violates the Lorentz symmetry, it does not reproduce some of the zero-temperature physics, such as the Silver-Blaze property.
On the other hand, we numerically obtain the critical exponents that are consistent with the preceding studies.
This consistency ensures the validity of our analysis, including the Taylor expansion method.

We have also analytically examined the flow equation with a different regulator involving the Matsubara frequency.
This examination agrees with our numerical results including the consistency with the MW theorem.
Besides, an analytic solution of a flow equation implies that the flatness around potential minimum demands higher numerical accuracy.
To precisely handle this shape of the potential, it is necessary to develop other numerical scheme beyond the Taylor expansion method and the Grid method.

This work provides the foundation in the nonperturbative analysis of pion BECs inside magnetars~\cite{Kaspi:2017fwg,Chatterjee:2018prm}.
In magnetars, a magnetic field would be strong enough to effectively realize one-dimensional space through the shrinking of the cyclotron radius of charged pions.
Also magnetars would have a large isospin chemical potential, which can be regarded as $\mu$ in this study, due to huge imbalance between proton and neutron numbers.
Since the FRG analysis for such a large $\mu$ and a low $d=1$ faces the numerical instability, we need to develop alternative approaches beyond the current methods.
Exploring the effect of strong magnetic field on the pion BECs, especially in relation to the MW theorem, is a promising direction for future work.

Another intriguing direction is to study the Berezinskii-Kosterlitz–Thouless (BKT) transition~\cite{Berezinsky:1970fr,Berezinsky:1972rfj,Kosterlitz:1973xp} in $d=2$ using FRG~\cite{Jakubczyk:2014isa,PhysRevA.85.063607}.
In nonrelativistic systems, the BKT transition temperature is closely tied to the chemical potential~\cite{PhysRevA.85.063607}.
The relativistic counterpart has not been fully explored, and our approaches could provide an informative basis for its analysis.

%%%%%%%%%%%%%%%%%%%%%%%%%%%%%%%%%%%%%%%%%%%%%%%%
%%%%%%%%%%%%%%%%%%%%%%%%%%%%%%%%%%%%%%%%%%%%%%%%

\section*{Acknowledgments}

K.~M. is supported by the Japan Society for the Promotion of Science (JSPS) KAKENHI under Grant No.~24K17052.
F.~T. acknowledges support from the Moritani Scholarship Foundation (No.~22051) and JST SPRING under Grant No.~JPMJSP2151.

%%%%%%%%%%%%%%%%%%%%%%%%%%%%%%%%%%%%%%%%%%%%%%%%
%%%%%%%%%%%%%%%%%%%%%%%%%%%%%%%%%%%%%%%%%%%%%%%%

%--- Bibliography ---%
\bibliography{paper}
%--- Bibliography ---%

%%%%%%%%%%%%%%%%%%%%%%%%%%%%%%%%%%%%%%%%%%%%%%%%
%%%%%%%%%%%%%%%%%%%%%%%%%%%%%%%%%%%%%%%%%%%%%%%%

\end{document}